\documentclass[5p,twocolumn,11pt,authoryear]{elsarticle}

\usepackage{amsmath}
\usepackage{amssymb}
\usepackage[utf8]{inputenc}
\usepackage[dvipsnames]{xcolor}
\usepackage{tabularx}
\newcolumntype{K}[1]{>{\centering\arraybackslash}p{#1}}
\newcolumntype{L}[1]{>{\arraybackslash}p{#1}}
\usepackage{graphicx}
\usepackage{tikz}
\usepackage[hidelinks,colorlinks=true,linkcolor=blue,citecolor=blue]{hyperref}

\makeatletter
\def\ps@pprintTitle{%
	\let\@oddhead\@empty
	\let\@evenhead\@empty
	\def\@oddfoot{\reset@font\hfil\thepage\hfil}
	\let\@evenfoot\@oddfoot
}
\makeatother

\begin{document}

\begin{abstract}
Nano- and microplastics are a growing threat for the environment, especially in aqueous habitats. For assessing the influence on the ecosystem and possible solution strategies, it is necessary to investigate the ``fate'' of microplastics in the environment. Microplastics are typically surrounded by natural organic matter, which can cause them to aggregate due to favorable interactions. However, the effect of microplastics' shape and flow conditions on this heteroaggregation is not well understood. To address this gap, we perform molecular dynamics simulations of heteroaggregation of different microplastic particle shapes with smaller spherical organic matter. We demonstrate that the shape had a strong impact on the aggregate structure. Microplastics with mostly smooth surfaces (e.g., spheres, rounded cubes) formed compact structures with a large number of neighbors with weak connection strength and a higher fractal dimension. Microplastics with edges and corners (e.g., cubes, plates) aggregated into more fractal structures with fewer neighbors, but with stronger connections. 
Using multiparticle collision dynamics, we investigated the behavior of aggregates under shear flow. The critical shear rate at which the aggregates break up is much larger for spherical and rounded cube microplastics, i.e, the compact aggregate structure of spheres outweighs their weaker connection strength.  Most notably, the rounded cube exhibited unexpectedly high resistance against breakup under shear. We attribute this to being fairly compact due to weaker, flexible neighbor connections, which are still strong enough to prevent microplastic particles to break off during shear flow.  Irrespective of the stronger connections between neighbouring microplastics, the fractal aggregates of cubes break up at lower shear rates. During shear, aggregates of all shapes can restructure. We find that cube aggregates reduced their radius of gyration significantly, indicating restructuring, while most neighbor connections were kept intact. Aggregates of spheres, however, kept their overall size while undergoing local rearrangements, that broke a significant portion of their neighbor interactions.
\end{abstract}

\title{Influence of Shape on Heteroaggregation of Model Microplastics: A Simulation Study}

\author[1]{B. Ru\c{s}en Argun}

\author[2]{Antonia Statt\fnref{fn1}}

\fntext[fn1]{Corresponding Author, \textit{Email:} \texttt{statt@illinois.edu}}

\affiliation[1]{organization={Mechanical Engineering, 
                              Grainger College of Engineering, University of Illinois},
                city={Urbana-Champaign},
                postcode={61801}, 
                state={IL},
                country={United States}}

\affiliation[2]{organization={Materials Science and Engineering, 
                              Grainger College of Engineering, University of Illinois},
                city={Urbana-Champaign},
                postcode={61801}, 
                state={IL},
                country={United States}}


\maketitle

\section{Introduction}

The production and use of plastics has exploded in recent decades~\citep{alimi2018microplastics}.
The resulting macroscopic plastic waste accumulates in  landfills and the environment, which continues to degrade into micro- and nanoscopic particles (MP/NP) through natural processes like mechanical, thermal, microbial degradation, photodegradation, and biochemical mechanisms. 
Additionally, micro- and nano-sized plastic particles can be discharged directly as byproduct of laundry or as additive in personal care products, causing an alarming and growing micro- and nanoplastic pollution in the environment~\citep{ryan2009monitoring}, especially in bodies of water~\citep{andrady2011microplastics}, but also in soil~\citep{guo2020source}, air ~\citep{gasperi2018microplastics}, and even in human blood ~\citep{leslie2022discovery}. In fact, plastics pollution is now abundant~\citep{Hamid2018,AKANYANGE2022113}.

This omnipresent pollution has been shown to be especially threatening for marine ecological systems \citep{besseling2019quantifying,triebskorn2019relevance}.
Nanoplastics can be toxic by themselves \citep{mattsson2018nanoplastics}, or they can function as carriers for toxic substances into living organisms \citep{velzeboer2014strong}.  
In an aquatic ecosystem, their transport mechanisms, whether plastic particles are suspended, aggregated, or deposited at the bottom are key points of interest to deepen our understanding about plastics pollution and the resulting implications for the environment and human health. 
These key points depend strongly on the size at which plastic particle clusters are found in the environment, which in turn, is determined by their aggregation behavior.
Therefore, a significant portion of research efforts have been focusing on the aggregation to determine the fate of micro- and nanoplastics in marine and freshwater environments \citep{wang2021review}. 

The concentration of micro- and nanoplastics in aqueous environments is commonly small in comparison to the concentration of other naturally occurring colloids, such as suspended particulate matter (SPM), natural organic matter (NOM) etc.~\citep{wagner2019things}.
Those natural colloids can facilitate or hinder the aggregation of micro-plastics due to their dominant concentration in environment. 
Hence, heteroaggregation (i.e., aggregation of particles of different species) is the prominent mechanism, rather than homoaggregation (i.e., aggregation of particles of same species)  to determine the fate of microplastics in aqueous environment~\citep{wang2015heteroaggregation}. 
NOMs have been shown to sterically stabilize the MPs in various studies \citep{cai2018effects, li2019interactions, singh2019understanding, yu2019aggregation, shams2020aggregation, wang2021aggregation}, and interestingly, they also have been shown to destabilize MPs through bridging, depending on the specifics of the system considered~\citep{cai2018effects}.   

Several experimental challenges complicate the study of aggregation behaviour of MP/NP in aquatic ecosystems.
First, it is difficult to separate and characterize MP/NP from environmental sources~\citep{fu2020separation}, the bulk of existing prior work used manufactured MP/NP either directly or after applying aging processes to make them more realistic \citep{yu2019aggregation}. Spherical and mostly identical model microplastics might not accurately reflect the  behavior of the microplastics found in nature. 
Second, the experimental parameter space that needs to be considered is enormous. 
For example, 
the effect of chemistry/surface functionalization \citep{tallec2019surface,shams2020aggregation}, size of the plastic particles \citep{sun2021difference}, polydispersity of the particles \citep{yu2019aggregation}, ionic strength, pH, salt type \citep{cai2018effects} all play a role in aggregation behavior.

Numerical models and simulations of microplastics can complement experimental investigations.   
Large-scale models simulating the transport of MPs inspired by previously developed engineered nanoparticle (ENP) fate models have been employed. \cite{besseling2017fate} used a hydrological model to simulate the fate and transport of MP in river systems. 
Similarly, \cite{domercq2022full} developed a mass-balance transport model for MPs in various aqueous conditions, and \cite{bigdeli2022lagrangian} provides a review of several particle tracking modelling methods used for MPs in waters. 
These models aim to determine the size distribution of MPs at various environmental conditions, and provide insight on the distribution and concentration of MPs for ecological studies. 
However, these transport models alone can not accurately account for heteroaggregation, biofouling, and defragmentation, as these phenomena depend on every single detail like chemistry, morphology of MPs, NOMs, and solvent conditions. These effects must be resolved on much shorter length and time scales. 

Colloidal simulations provide the opportunity to systematically study single effects caused by the many parameters involved in the MPs heteroaggregation. Thus, these particle based simulations can give further insights that can be challenging to resolve in experiments. 
The insights are also needed for accurate large-scale transport models. 
However, there are only few publications investigating MPs heteroaggregation from a colloidal simulation perspective.
\cite{clavier2019determination} used Monte Carlo (MC) simulations investigate heteroaggregation of potentially hazardous engineered nanoparticles with natural organic matter in aqueous conditions. 
They've shown that natural organic matter can bridge microplastics into large heteroaggregates or stabilize them depending on the initial concentrations and solvent conditions. 
A statistical thermodynamic heteroaggregation model was employed in \cite{wheeler2021statistical}, showing that nano-micro contaminants under environmentally relevant conditions don't heteroaggregate to the completion even with favorable bonding energies and thus stay mostly suspended.
\cite{wang2022effects} used Molecular Dynamics (MD) to simulate polyethylene chains with lipid bilayers in order to understand the behaviour of nanoplastics in organisms. 
\cite{li2021adsorption} studied pesticide adsorption onto polyethylene microplastics in water environment using atomistic MD simulations in addition to experiments.

Due to experimental challenges, most work to date focused on the investigation of spherical model micro- and nanoplastic particles. However, irregular shapes are most commonly found in environmental systems~\citep{wang2021review}.
Most notably, \cite{pradel2021stabilization} and \cite{dong2021aggregation} have shown that non-spherical morphologies favor aggregation.
\cite{isachenko2020catching} employed a stochastic simulation to calculate terminal velocity of MPs in the water column and found that it is mostly affected by the density and size of MPs rather than their shape. 

Simulations of non-spherical colloidal of particles were performed by several authors, \cite{kobayashi2022self} modeled cubes as group of spherical beads, similarly, \cite{nguyen2019aspherical} used groups of beads to simulate irregularly shaped particles in MD.  Alternatively, self-assembly of several different geometries, including rods and plates, was investigated using patchy particles~\citep{zhang2004self}.
The effect of shape on phase behavior was also studied by finely tuning geometries of athermal colloids using Monte Carlo simulations \citep{marechal2010phase,ni2012phase,avvisati2015self}. Most prior simulation work focused on equilibrium behavior of shapes.   

Realistic conditions are not necessarily quiescent, currents are abundant in aqueous environments. However, restructuring and breakup effect of aggregates due to flow is rarely considered explicitly in existing literature. 
Shear flow, for example, has a considerable impact on behaviour of colloidal systems. It has been shown that weakly bonded NP/MP aggregates can breakup under shear flow under environmental conditions~\citep{enfrin2020release}.
Shear flow does not only break aggregates, but can restructure them, as shown by \cite{becker2009restructuring} using Discrete Element Method (DEM) simulations. 
They investigated the balance between aggregation and breakup, which determined the aggregate size  distribution~\citep{oles1992shear}. To our knowledge, prior studies focused mostly on spherical model particles. 


Integrating shear flow into colloidal simulations to understand its effect on aggregation can be achieved with several methods, e.g.~\cite{frungieri2020shear} used MC and DEM to simulate heteroaggregation under shear flow. 
This paper confirmed that strong asymmetry in initial monomer concentration of the two species results in size stabilization of heteroaggregates.~\cite{harshe2012breakage} simulated breakup of aggregates obtained by MC using Stokesian Dynamics.
They were able to calculate breakage rates that can be used as kernels for population balance equations.

Our work investigates the structure of heteroaggregates formed by NOMs and different microplastic (MP) shapes. First, we describe the methods and models used to simulate heteroaggregation in Section \ref{sec:model}. Resulting clusters and their structure is discussed in Section \ref{sec:results_eq}.
We then studied the stability of heteroaggregates under shear flow in Section \ref{sec:results_flow}. Overall, our work addresses two important, but often overlooked effects in nanopastic aggregation,i.e.\ shape of plastic particles and resulting stability under shear. We conclude with a discussion in Section \ref{sec:conclusion}.

\section{Model and Methods\label{sec:model}}

To address the influence of shape on heteroaggregation, eight different microplastic (MP) particle shapes were modeled as groups of smaller constituent beads, as listed in Table~\ref{tab:shapes}. 
Microplastics in the environment may have irregular and asymmetric morphologies~\citep{wang2021review} that do not correspond to any of the fairly regular MP shapes used in this study. 
Instead of attempting to model all possible shapes, this work focused on systematic variations that might be helpful for understanding the effect of edges, corners, or smooth surfaces of MPs on their heteroaggregation behavior.  

Natural organic matter (NOM) particles were modeled as single spherical beads of the same size as the constituent beads of the MP shapes. 
MP shapes were held rigid during the simulation using a rigidity constraint~\citep{nguyen2011rigid,glaser2020pressure}.
The unit of length and mass were set by the individual beads of MPs and NOMs, which had size $1\sigma$ and mass $M=1m$.  

The surface area $A_\textrm{center}$ and volume $V_\textrm{center}$ of a shape was calculated with respect to the dimensions given by the center locations of the constituent beads. However, the effective area $A_\textrm{eff}$ and effective volume $V_\textrm{eff}$ are more indicative for the aggregation and shear flow behavior, as estimation for the hydrodynamic radius. Both $A_\textrm{eff}$ and $V_\textrm{eff}$ were calculated by including the constituent bead excluded volume by adding $0.3\sigma$ to each direction.
All eight MP shapes had similar effective volumes $V_\textrm{eff}$ of approx. $100\sigma^3$, resulting in different surface areas $A_\textrm{eff}$, as stated in table~\ref{tab:shapes}. The small differences were caused by the limited resolution given by the constituent bead size. 
 
Four shapes were cylindrical, with elliptic cross-sections of decreasing eccentricity; 0.88 for ellipse 1, 0.68 for ellipse 2, and 0, which corresponds to a regular cylinder with circular cross section. 
We also simulated a thinner and taller cylinder, which was denoted as rod. 
Additionally, cube, plate, rounded cube, and sphere MP shapes were modeled. 
These eight shapes represent archetypical geometries that allowed for systematic investigation of the effect of shape,  edges, corners, and smooth surfaces.

\begin{table*}[ht!]
\renewcommand{\arraystretch}{1.2}
\small
  \caption{Microplastics shapes, their surface area $A_\textrm{center}$ and volume $V_\textrm{center}$ as calculated with respect to the dimensions given by the center of the constituent beads, number of beads in each shape, and their dimensions. The effective area $A_\textrm{eff}$ and effective volume $V_\textrm{eff}$ are also given.}
  \label{tab:shapes}
  \begin{tabular}{L{1.7cm}K{1.5cm}K{1.5cm}K{1.5cm}K{1.65cm}K{1.5cm}K{1.5cm}K{1.88cm}K{1.5cm}}
    \hline
    & Cube & Plate & Rod & Ellipse 1 & Ellipse 2 & Cylinder & Rounded Cube & Sphere \\[1.2em]
    \hline
   $\#$ of beads & 152 & 138 & 128 & 123 & 118 & 120 &  140 & 100 \\
   Dimensions (center) & $a=4.15\sigma$ & $a=1.8\sigma$, $b=5.4\sigma$, $c=6.3\sigma$ & $L=9.9\sigma$, $r=1.45\sigma$ & $L=5.4\sigma$, $a=2.85\sigma$, $b=1.35\sigma$ &  $L=5.4\sigma$, $a=2.3\sigma$, $b=1.68\sigma$ &  $L=5.4\sigma$, $R=2\sigma$ & $a=4.35\sigma$, $r_\mathrm{curv}=0.93\sigma$ & $R=2.6\sigma$ \\
   $V_\textrm{center}\,[\sigma^3]$ & 71.5 & 61.2 & 65.4 & 65.3 & 65.6 & 67.9 & 74.1 & 73.6 \\
    $V_\textrm{eff}\,[\sigma^3]$ & 107.2 & 99.4 & 101.0 & 97.9 & 97.0 & 99.7 & 104.9 & 102.2 \\
  $ A_\textrm{center}\,[\sigma^2]$ & 103.3 & 110.6 & 103.4 &  97.7 & 92.2 & 93.3 & 84.7 & 73.6 \\
    $ A_\textrm{eff}\,[\sigma^2]$ & 135.4 & 145 & 137 &  125.3 & 120 & 120 & 110 & 105.6 \\
    \hline
  \end{tabular}
   \hspace*{2.0cm}\includegraphics[width=0.88\textwidth]{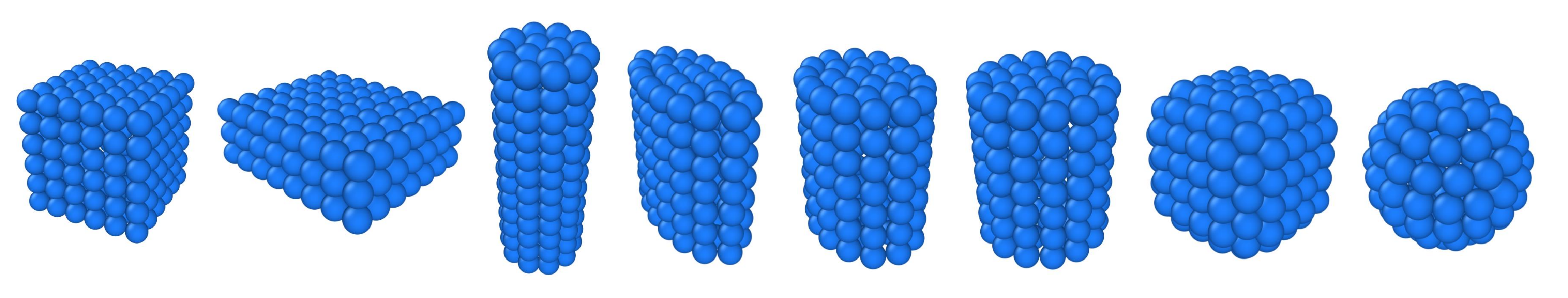}
\end{table*}


\begin{figure}
    \centering
    \includegraphics[width=8cm]{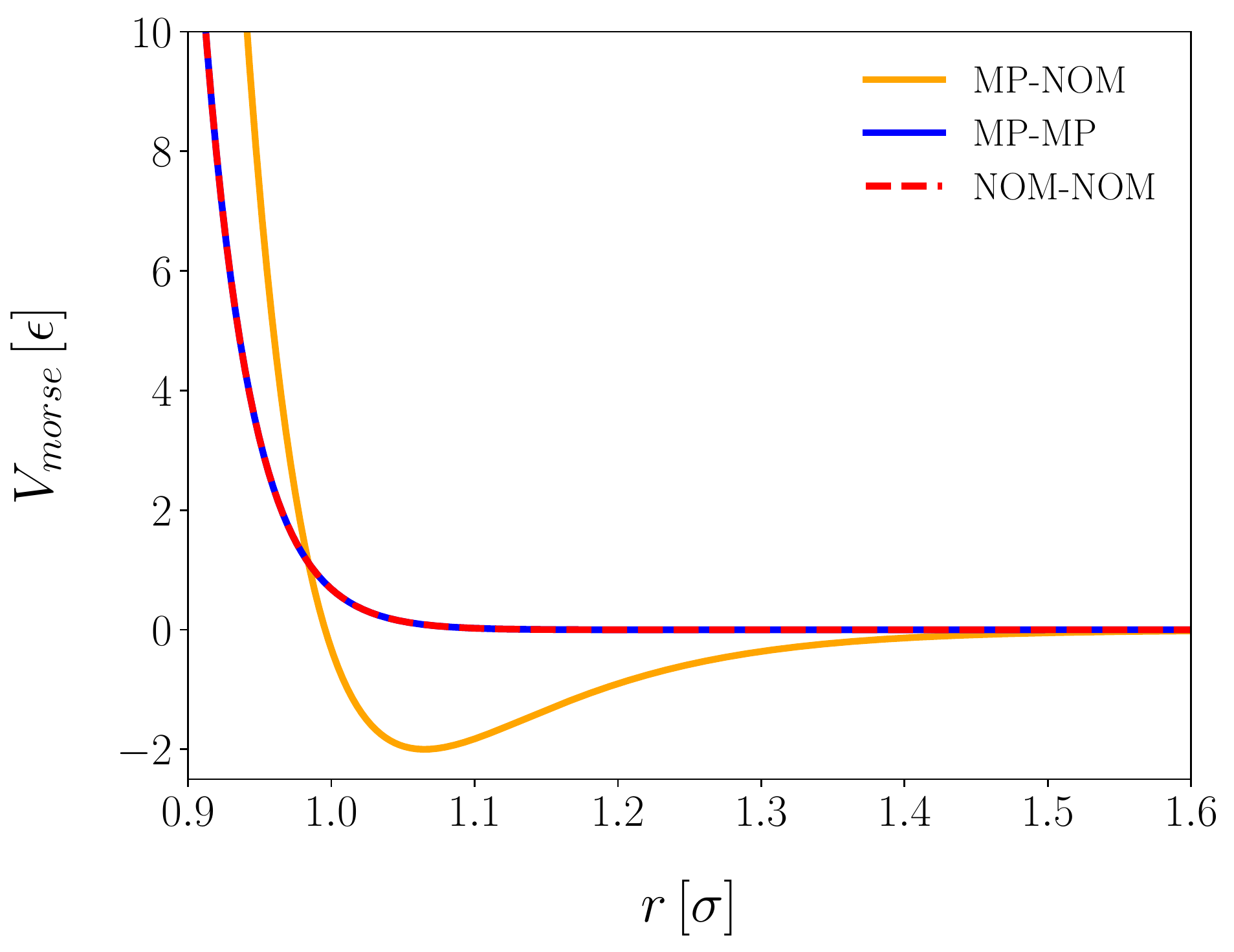}  
    \caption{Morse potential for MP constituent bead and NOM interactions. The parameters were $D_0 = 0.001 \epsilon$, $  \alpha=15\sigma^{-1}$, and $r_0 = 1.22 \sigma $ for MP-MP and NOM-NOM interactions and $D_0=2\epsilon$, $\alpha=10\sigma^{-1}$, and $r_0=1.065\sigma$ for MP-NOM interaction. }
    \label{fig:morse_potential}
\end{figure}

Each constitutive bead in a MP shape interacted separately with a Morse potential 
\begin{align*}
\label{eqn:morse}
V_{\mathrm{morse}}(r)  =\, & D_0 \left[ e^{-2\alpha\left(r-r_0\right)} -2e^{-\alpha\left(r-r_0\right)} \right] &      r < r_{\mathrm{cut}} \\
                       =\, & 0 \quad . & r \ge r_{\mathrm{cut}} 
\end{align*}
All pair interactions were cut at $r_{\mathrm{cut}}=1.75\sigma$, and are shown in Fig.~\ref{fig:morse_potential}. The unit of energy was chosen to be $\epsilon$. 
The Morse potential was used with $D_0 = 0.001 \epsilon$, $  \alpha=15\sigma^{-1}$ and $r_0 = 1.22 \sigma $, to model the repulsive interactions between same species, i.e., NOM-NOM and MP-MP. 
A short range attraction between MP constituent beads and NOM beads was modeled with $\alpha=10\sigma^{-1}$, where attraction strength at $r_0=1.065\sigma$ was given by $D_0=2\epsilon$. 
The choice of parameters for all pair interactions led to a similar excluded volume, when considering the repulsive part at $r\leq 1\sigma$. 
This set of pair interactions promoted heteroaggregation as the single aggregation mechanism in the system, and no homoaggregation between just MPs or just NOMs was observed in any simulation.


\subsection{Quiescent Aggregation \label{sec:model_equilibrium}}

Heteroaggregation of MPs with NOMs was simulated using the open-source molecular dynamics (MD) software package HOOMD-blue v2.9.4 \citep{Howard2018,Howard2019,anderson2020hoomd}.
A time step of 0.005 $\tau$ was used,  where $\tau = \sigma\sqrt{m/\epsilon}$ was the unit of time. A constant temperature of $T=1 \epsilon/k_B$ was maintained using a Langevin thermostat with a friction coefficient  of 1 $m/\tau$ applied to all beads.  
Instead of strong inter-particle bonds, a rigidity constraint was employed to keep MP shapes undeformed throughout the simulation \citep{nguyen2011rigid,glaser2020pressure} for computational efficiency. 

It was not computationally feasible to perform  MD simulations at the environmentally relevant low MP concentrations \citep{eriksen2013microplastic}.
To mimic the low concentrations of MPs as found in aqueous  environments, MPs-NOMs were added sequentially to the box, while the MPs were allowed to aggregate during the simulation. 
Simulations were initialized with 2 MPs and 250 NOMs in a cubic box with side length $L = 20 \sigma$. 
After each cycle of $1000\tau$, the clustering algorithm DBSCAN \citep{ester1996density} (with a neighbor distance cutoff of $1.5 \sigma$ and minimum number of 2 for core points) was used to to check whether all the of MPs in the box had aggregated into a single cluster.
If they were clustered, an additional MP shape and 125 NOMs were added into the box randomly without overlap, and the next cycle of $1000\tau$ started. If the MPs were not clustered, no particles were added and the simulation was continued for another $1000\tau$ before checking again. 
During this sequential aggregation scheme, the number density of MPs ($\rho_{MP}=0.00025\sigma^{-3}$) and NOMs ($\rho_{NOM}=0.03125\sigma^{-3}$) were kept constant by appropriate increases of the box size as new particles were added to the box. The aggregation was continued until each cluster reached a size of 70 MPs, which was large enough to investigate differences in structure and stability. Overall, 10 heteroaggregates were obtained independently to increase statistics for each of the shapes listed in Table~\ref{tab:shapes}.

As an alternative approach, Monte Carlo (MC) Simulations with cluster moves were used to obtain aggregates. 
The resulting clusters were more fractal, less compact, and had less bonded neighbors.
This indicated that they were not efficiently equilibrated, despite the use of several highly efficient cluster moves~\citep{Whitelam2007,sinkovits2012rejection}. 
Additionally, straightforward MD simulations that contained all shapes from the beginning at random initial conditions were performed. Similarly to the MC simulations, the aggregates obtained could not be equilibrated within a reasonable simulation time, especially for shapes with pronounced edges and corners, like cubes. 

Aggregate structure was evaluated with two quantities: the average number of bonded neighboring MP shapes, or bonded neighbor number (BNN) and the average connection strength (ACS).
BNN was defined as the average number of MPs that each MP in an aggregate was bonded to by at least one NOM (i.e., $\#$ of bonded MP neighbors, where a bond was an NOM that was in attraction with both of MPs). For this calculation, the bonding cutoff was 1.4$\sigma$, where the NOM-MP attraction is close to zero (see Fig.~\ref{fig:morse_potential}). 
Higher BNN indicated a more compact aggregate with more nearest neighbors, indirectly indicating the fractal dimension of the aggregate. 
The average connection strength (ACS) was given by the average number of NOMs that were bridging two neighboring MPs in a aggregate.  
ACS served as a measure of effective attraction/bond strength between neighboring MPs of an aggregate.

\subsection{Aggregates Under Shear Flow}

In conditions relevant to the environment, the MPs aggregates could be subjected to flow, which may break them into smaller fragments. 
To study these effects systematically, shear flow with a constant velocity gradient between two walls was used, where the top wall moves in $+x$ and the bottom wall in $-x$. 
Multi-particle collision dynamics (MPCD)\citep{malevanets1999mesoscopic}, a standard mesoscale simulation method, was used to incorporate a background fluid and hydrodynamic interactions into the simulation. We refer the reader to \citep{ripoll2005dynamic,gompper2009multi,poblete2014hydrodynamics,howard2019modeling} for details on MPCD. 
MPCD was used with the stochastic rotation dynamics (SRD) collision rule~\citep{malevanets1999mesoscopic}, as implemented in HOOMD 2.9.4 \citep{howard2018efficient}.
MPCD treats the solvent explicitly as point-like particles with unit mass $m$. The number density of solvent particles $\rho$ was set to $5\sigma^{-3}$ and correspondingly, the mass of NOMs and MPs constituent beads was set to $M=5m$ to ensure proper solute-solvent coupling \citep{gompper2009multi}. Since no dynamical quantities were calculated from the quiescent cluster aggregation simulations, the change in mass and simulation method did not cause any inconsistencies. 
The solvent particles in MPCD were simulated in two steps: a streaming step where they moved ballistically and a collision step, where they were also coupled with MD particles (i.e., MPs and NOMs). The collision period was set to $20$, i.e., $\lambda = 0.02\tau$, and the collision angle was set to $130^\circ$.
MD particles were simulated using the standard velocity-Verlet algorithm with $\Delta t = 0.001 \tau$ and a tree neighbor list~\citep{Howard2019}, and they interacted with the solvent beads only during the collision step.

A theoretical prediction of the Schmidt number $Sc$ can be calculated for MPCD solvents, since the viscosity can be determined readily~\citep{gompper2009multi}.
For the parameters used in this study, the Schmidt number was $\sim250$ which was sufficient to approximately reproduce the properties of water-like fluids, which typically have $Sc =10^2 - 10^3$ \citep{gompper2009multi}.

To create the linear shear flow profile, the periodic boundaries in $+z$ and $-z$ directions were replaced with repulsive walls. 
These walls moved in $\pm x$ direction with constant speed during the simulation, which was varied between 0.003 and 0.1 $\sigma\tau^{-1}$ to create different shear rates.
A constant fluid velocity gradient in $z$ direction was established by imposing a no-slip bounce back collision rule for the solvent particles colliding with a wall.  
We also used virtual filler with MPCD particles at a density of $5\sigma^{-3}$ in the walls, which is known to improve the no-slip wall boundary condition \citep{Imperio2011,bolintineanu2012}.
It is important to note that the final box size was large enough to ensure that the clusters were sufficiently far away from the walls to prevent any interaction between aggregate and walls. 
Since the rigidity constraint that was used in the MD simulations of aggregation is not easily compatible with MPCD, a harmonic bond potential with a large stiffness of $k=4000 \epsilon/\sigma^{2}$ was used between neighboring MP beads of the same shape to ensure the rigidity of MP shapes \citep{wani2022diffusion}. 


\section{Results \label{sec:results}}

\subsection{Structure of Aggregates in Equilibrium  \label{sec:results_eq}} 

\begin{figure}
    \centering
    \includegraphics[width=8cm]{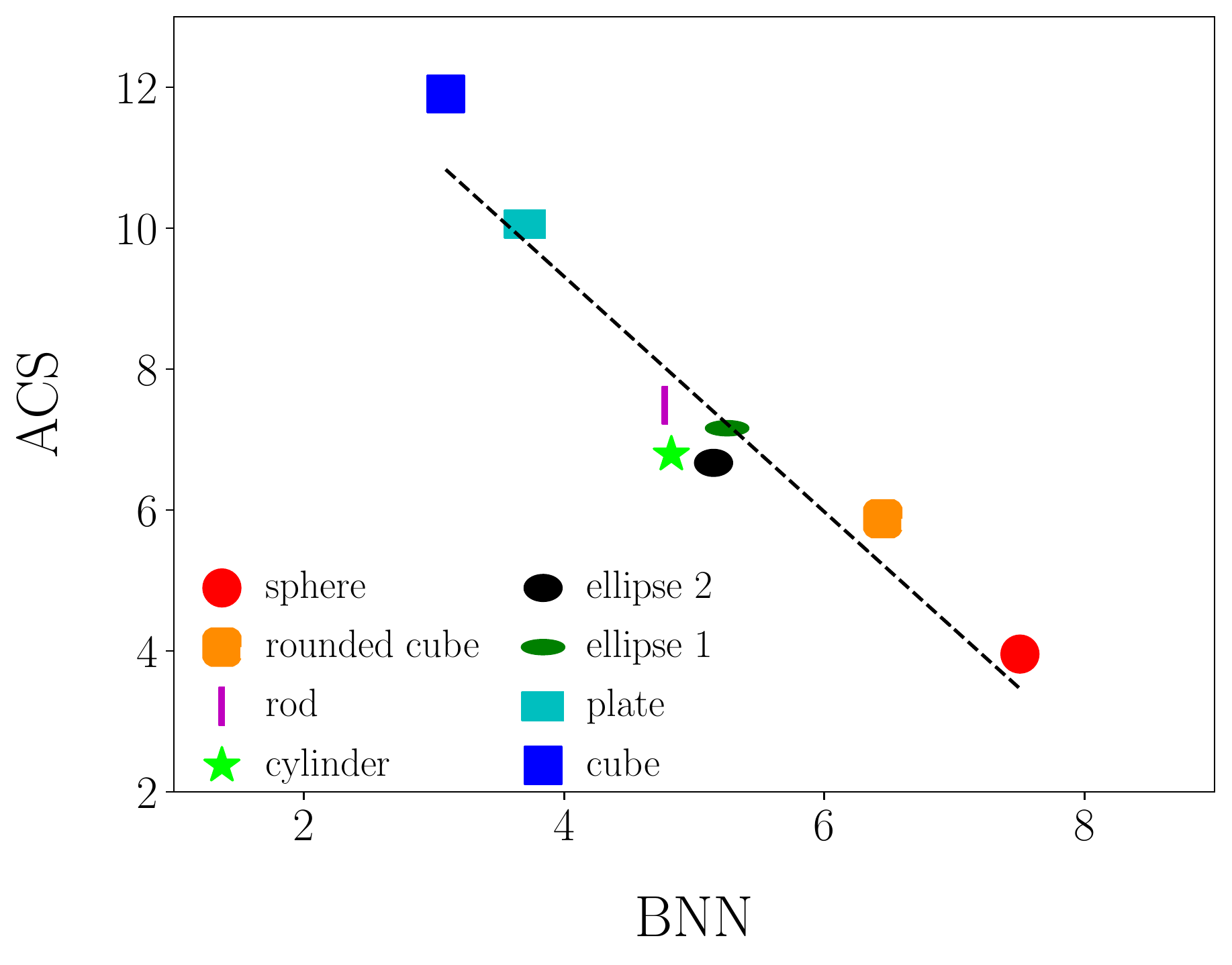}
    \caption{Average Connection Strength (ACS) as function of bonded nearest neighbors (BNN) for heteroaggregates of all different MP shapes (sphere, soft cube, rod, cylinder, ellipse 1 \& 2, plate, and cube) as indicated by the legend. The dashed line is a guide to the eye.}
    \label{fig:ad_vs_cf_v3}
\end{figure}
\begin{figure}[tph]
    \centering
    \includegraphics[width=8cm]{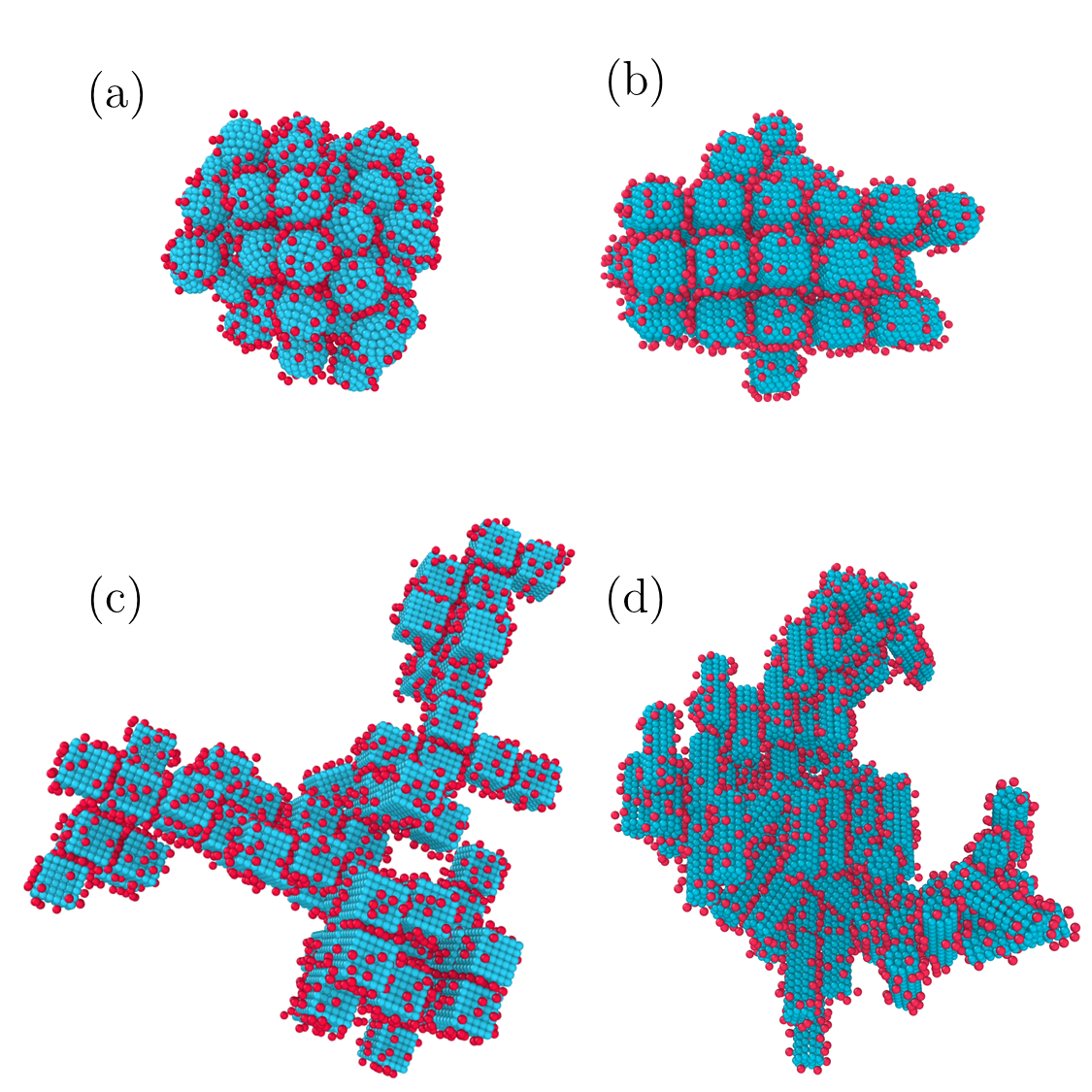}
    \caption{Snapshots of typical heteroaggregates of MP (a)  spheres, (b) rounded cubes, (c) cubes  and (d)  rods. MP constituent beads are shown in blue, NOMs are red.}
    \label{fig:snapshots_aggregates}
\end{figure}

First, we investigated the structure of aggregates in equilibrium as obtained by the simulations described in Section~\ref{sec:model_equilibrium}.
The average number of bonded neighbor MPs (BNN) and the average connection strength (ACS) between neighbors in the aggregates of different shapes were measured for each shape and plotted in Fig.~\ref{fig:ad_vs_cf_v3}, where the results were averaged over 10 heteroaggregates of each shape, with resulting error bars of the size of the symbols in the figure. 
The inverse correlation between the number of neighbours BNN and connection strength ACS is evident.
At one end, heteroaggregates of spherical MPs have the highest number of neighbors of around 8 and lowest connection strength of around 3.5, i.e, the MPs in a cluster have many loosely connected neighbors. 
At the other end,  heteroaggregates of cubes and plates have the highest connection strength of approx. 12 and lowest number of neighbors around 3, i.e, the MPs in a cluster have very few strongly connected neighbors.
Interestingly, the product of connection strength ACS and average number of neighbors BNN was approximately the same for all shapes, independently of shape. 
The geometry of cubes and plates allow MPs to be bridged across their flat faces, which can accommodate a higher number of NOMs compared to curved surface of spheres, leading to higher ACS.
Once two flat surfaces were bridged by a large number of NOMs, it was unlikely for that connection to be broken or rearranged during the rest of the simulation. 
Between two MPs that were connected across flat surfaces surrounded by edges and corners, an effective resistance against rotation and torsion was created, that effectively hindered the restructuring of aggregates of cubes and plates into more compact geometries. 
However, due to their curved surface, spheres usually do not have more than approx. 4 NOMs bridging them (see Fig.~\ref{fig:ad_vs_cf_v3}), thus allowing for easier restructuring into more compact aggregates, which also explains why they have a higher number of neighbors in an aggregate. Typical snapshots of sphere, rounded cube, cube and rod MP aggregates were shown in Fig.~\ref{fig:snapshots_aggregates}. 

To focus on the effect of edges and corners, we simulated rounded cube MPs which were obtained by smoothing out the edges and corners of a cube while keeping some amount of the flat faces intact. 
The resulting structure of rounded cube aggregates were similar to spheres. 
In fact, they were more similar to spheres than all other shapes investigated here, as shown in Fig.\ref{fig:ad_vs_cf_v3}, underscoring the importance of edges and corners, rather than just flat faces.  
The other four shapes (3 cylinders and rod) have an amount of edges and corners between the sphere and cubes, and as a result their BNN and ACS was found to be between spheres and cubes. 
The resulting heteroaggregates of those four shapes were very similar, a typical example for rod aggregates is shown in Fig.\ref{fig:snapshots_aggregates}(d). The effect of shape was most prominent when comparing substantially different geometries like spheres and cubes.

\begin{figure}[tph]
    \centering
    \includegraphics[width=8cm]{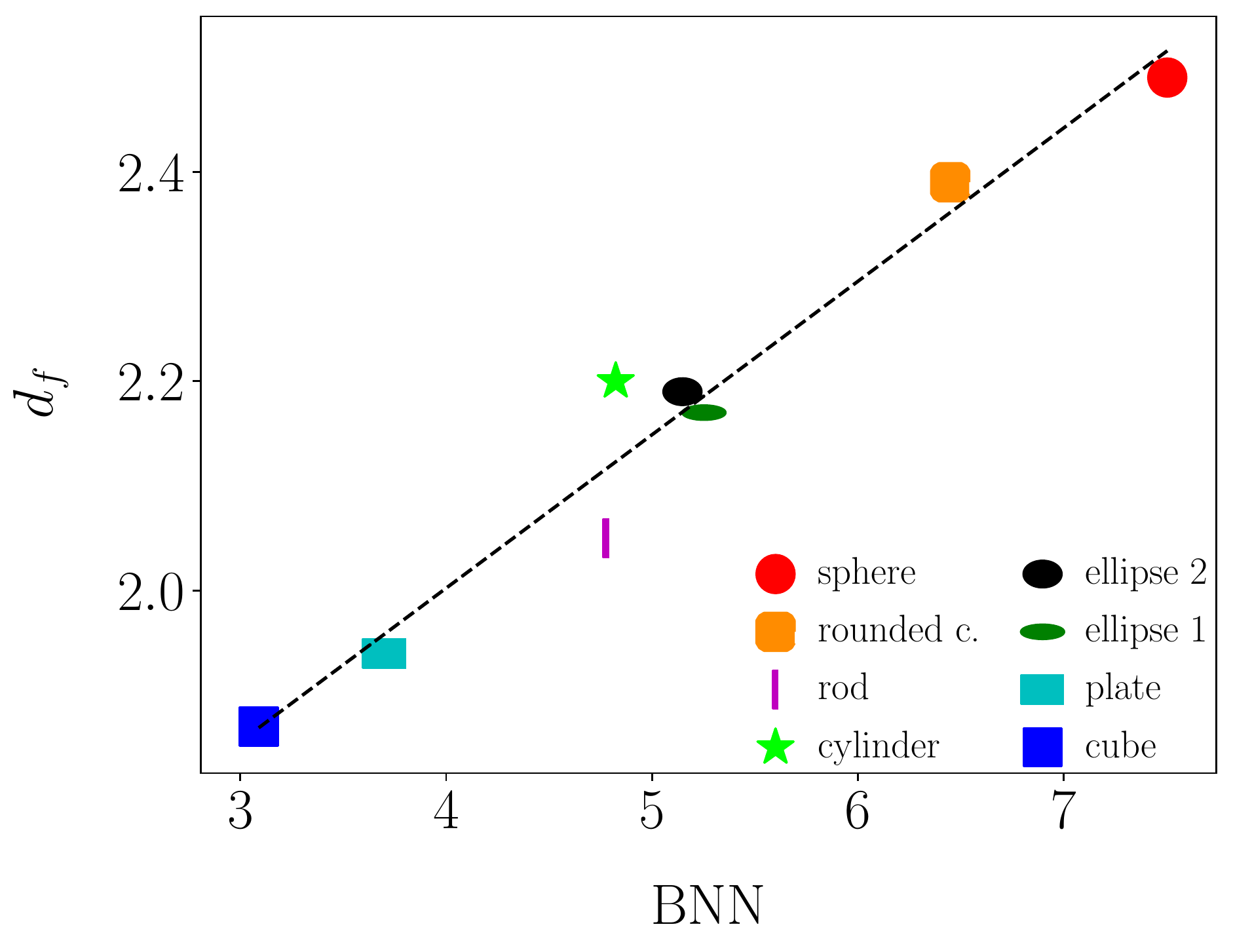}
    \caption{Fractal dimension $d_f$ of aggregates with different shapes (as indicated by the legend) as function of BNN. The dashed line is a guide to the eye.}
    \label{fig:df}
\end{figure}

A popular way to quantify the aggregate structures is their fractal dimension, $d_f$. 
Here, $d_f$ was defined as the exponent relating the size of the aggregate $R$ to the number of particles in the aggregate $N$
\begin{equation}
    N = k_0 (R/a)^{d_f} \quad ,
\end{equation}  
where $a=2.9\sigma$ is the effective size of the MP, i.e, the radius of a corresponding sphere with same volume. The parameter $k_0$ is a proportionality prefactor. Here, we set the prefactor $k_0$ to be unity for simplicity. Commonly, the value of $k_0$ is found to be close to one in three dimensions~\citep{sorensen1997prefactor}.

In this study, we found that the fractal dimension of an aggregate was linearly correlated with the average number of neighbors, as shown in Fig.~\ref{fig:df}.
The aggregates of cubes have the lowest $d_f$ of approx. 1.8, whereas spheres have a fractal dimension of approx 2.5. 
The low $d_f$ for cubes is close to values expected from  diffusion limited cluster aggregation (DLCA)~\citep{sorensen1997prefactor}, where  clusters modeled by random growth and attachment with no restructuring. 
The lack of restructuring in cube and plate heteroaggregates explains the observed fractal dimensions close to typical DLCA aggregates.

Rods also exhibited a lower $d_f$ of around 2.1, due to their asymmetric shape. 
The preferences of rod MP to align parallel in aggregates contributed to their lower fractal dimension. 
For aggregates as small as the ones studied here, with $N=70$, it is not straightforward to disentangle those two contributions or to fit the prefactor $k_0$ accurately.  
Restructuring in sphere and rounded cube heteroaggregates led to higher $d_f$ values.  

The simulations performed here did not model realistic aggregation kinetics directly.
However, the results presented in this work allow for some speculations.  
The aggregation rate should be proportional to both collision frequency and collision efficiency, and $d_f$ affects both those quantities~\citep{thill2001flocs}. 
Aggregates of lower $d_f$ will have larger effective sizes as they grow, leading to potentially more collisions with other aggregates. 
As we have shown here, MPs with flat surfaces (i.e., cubes and plates) had lower fractal dimensions, whereas spherical MP aggregates were compact and have much higher fractal dimensions. 
In addition to size considerations, the hydrodynamic repulsion forces can be reduced between two fractal aggregates as they come in contact, due to their open structure. This effect could increase collision efficiency, i.e.\, the fraction of aggregate collisions that lead to further aggregation. 
Thus, by increasing both frequency and efficiency of collision, a lower $d_f$ can have a compounding effect on aggregation rate. 
Therefore, the results presented here point towards faster aggregation of cubes and plates compared to rounded cubes and spheres, due to their difference in fractal dimensions.

\subsection{Aggregates under Shear Flow \label{sec:results_flow}}

In the previous section, structural properties of equilibrated clusters were investigated. 
Here, we study their stability under shear flow using MPCD.
A priori, it is not obvious whether a compact cluster with many loosely connected neighbors or a cluster with fewer, but strongly connected neighbours would be more stable under shear flow. 
To investigate this, a shear rate of $7.14 \cdot 10^{-4} \tau^{-1}$ was applied to all clusters of different MP shapes.

\begin{figure}
    \centering
    \includegraphics[width=8cm]{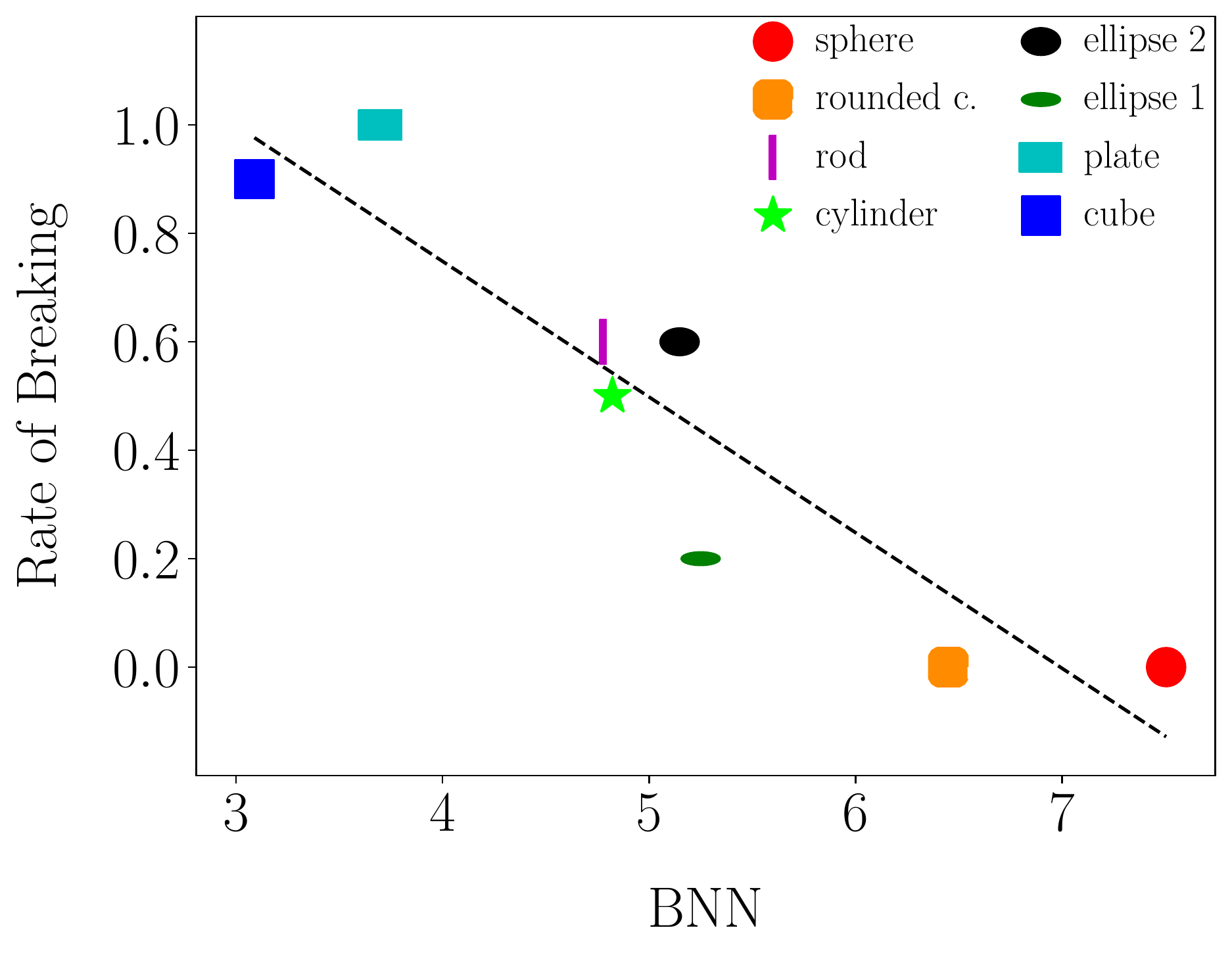}
    \caption{Rate of Breaking as function of number of bonded neighbours, BNN. The dashed line is a guide to the eye.}
    \label{fig:breaking_vs_ad}
\end{figure}

The rate of breaking, i.e., the fraction of the ten clusters of each MP shape that broke up into smaller clusters (where the largest cluster had less than 90\% of the initial cluster particles) was plotted in Fig.~\ref{fig:breaking_vs_ad} for all shapes.
At this particular shear rate, almost all cube and plate MPs heteroaggregates break up under shear. 
In contrast, aggregates of sphere and rounded cube MPs were resistant against the flow and did not show any breakup at this particular shear rate.  

Among the measures to quantify the structure of heteroaggregates, BNN correlated most with the resistance against breaking of an aggregate, as evident in Fig.~\ref{fig:breaking_vs_ad}.
With this data, it is clear that aggregates with higher number of neighbors broke up less. 
On the other hand, a similar observation cannot be made for connection strength. 
Clusters of cubes and plates have higher effective connection strength among their MPs (see Fig.\ref{fig:ad_vs_cf_v3}), however, they broke up more frequently. 
The rate of breaking is also not correlated with the aspect ratio of the shape, as evident by comparing rods and  cylinders. 
Overall, we conclude that a higher number of neighbors provided more stability against shear breakup than few strong connections between neighbors did.

\begin{figure}
    \centering
    \includegraphics[width=8cm]{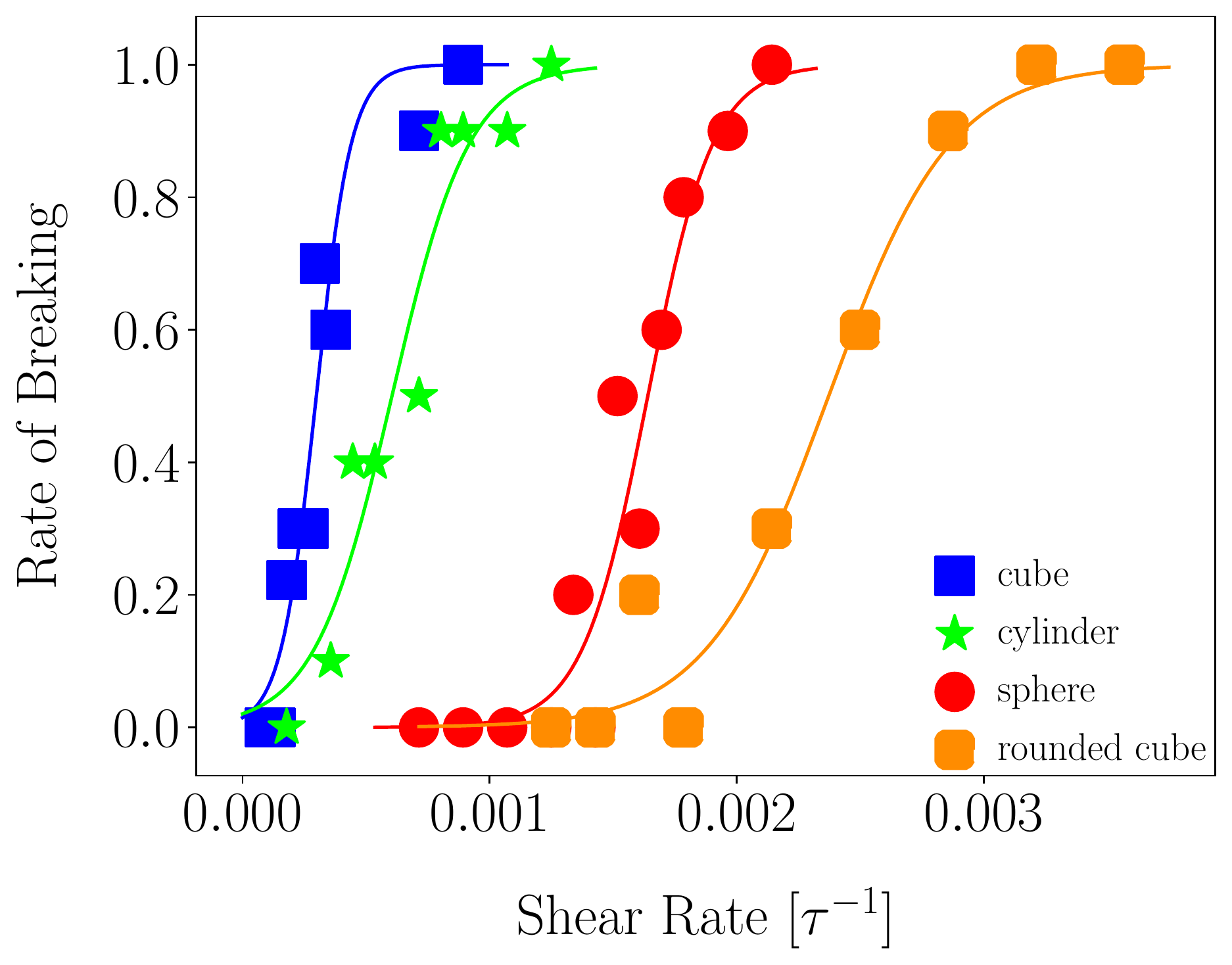}
    \caption{Rate of breaking as function of shear rate for different shapes, as indicated in the legend. The solid lines are tanh fits.}
    \label{fig:critical_shear_rate}
\end{figure}

To investigate their behavior under shear further, a limited number of shapes were investigated with varying shear rates. Aggregates of cubes, cylinders, rounded cubes, and spheres were subjected to a range of shear rates, since they were representing  archetypal clusters: open, fractal cluster with few, strongly connected neighbors, compact clusters with many weakly connected neighbors, and cylinders in between, with a highly asymmetric MP shape. 
The goal was to determine the critical shear rate at which each cluster started breaking. 

The shape of the MP also has a significant effect on the critical shear rate of the heteroaggregates of size $70$ MPs.
The critical shear rates to break 50\% of all clusters were obtained from tangent hyperbolic fits
\begin{equation*}
    f(s) = \frac{1}{2}\left[\tanh\left(\frac{s-c_s}{d}\right)+1\right] \quad,
\end{equation*}
resulting in critical shear rates of $c_s =3 \times 10^{-4} \tau^{-1}$, $6\times 10^{-4}\tau^{-1}$,  $1.64\times 10^{-3}\tau^{-1}$, and  $2.36\times 10^{-3}\tau^{-1}$ , for cubes, cylinders,
spheres, and rounded cubes, respectively. As shown in Fig.~\ref{fig:critical_shear_rate} the shear rate that was needed to break aggregates of spherical MPs was approx. five times higher than the one for cubes. 
Interestingly, this critical shear rate is \textit{not} a simple linear function of either BNN, ACS, or $d_f$.  While the critical shear rate is overall positively correlated with BNN, the rounded cube exhibited unexpectedly stable behavior against breaking up under shear, leading to a much higher critical shear rate. We attribute this to a favorable combination of compactness (i.e., high $d_f \gtrsim 2.3$) due to weak flexible neighbor connections, i.e.\, BNN and ACS around 6.5, that are still strong enough to prevent MP breaking off during shear.

\begin{figure}
    \centering
    \includegraphics[width=8cm]{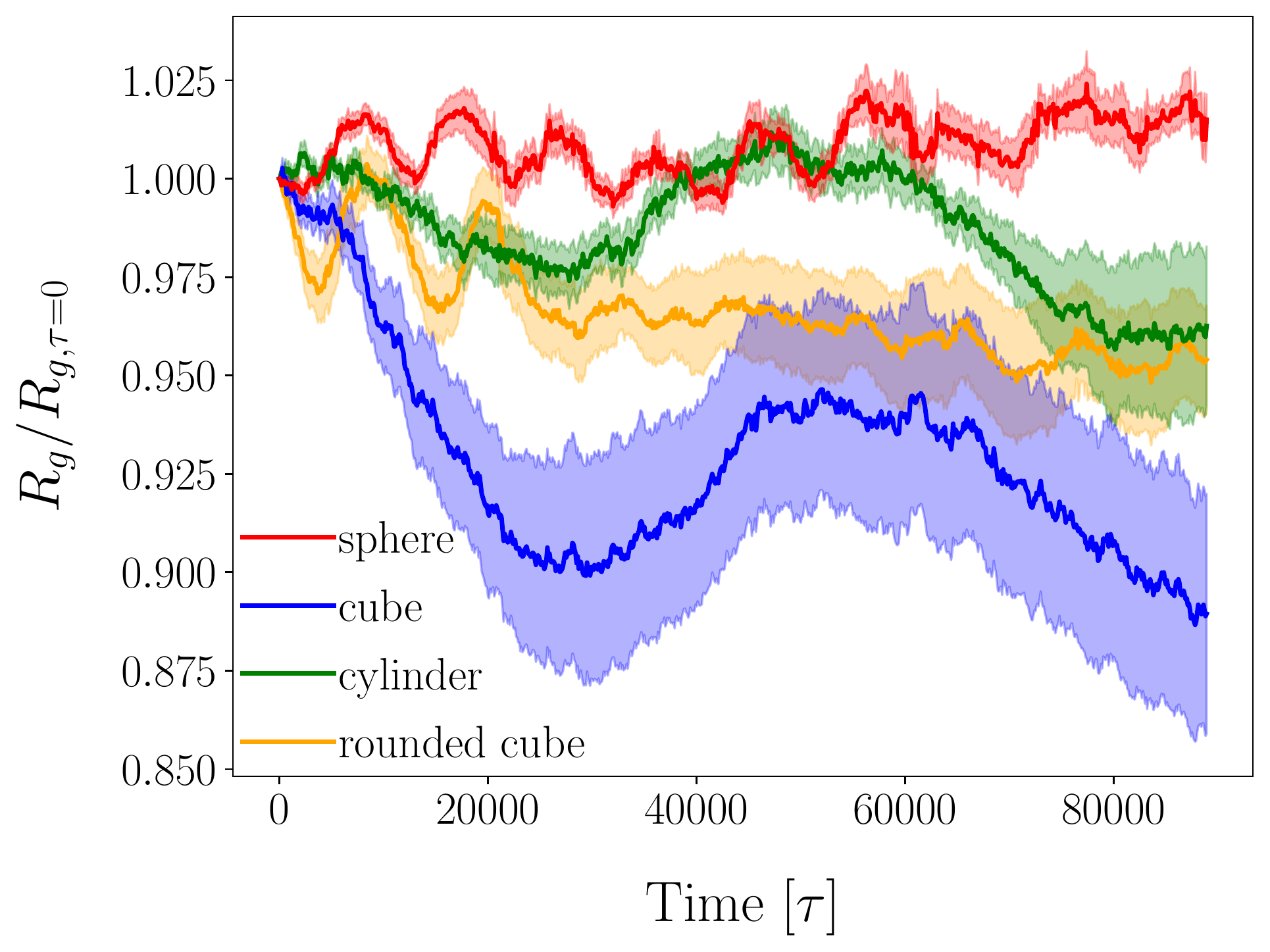}
    \caption{Radius of gyration $R_g$ as function of time at shear rates $1.36 \times 10^{-4} \tau^{-1}$ (cube and cylinder) and $7.85 \times 10^{-4} \tau^{-1}$ (sphere and rounded cube) averaged over 10 aggregates. The shaded area indicates the variance in the averaged data.}
    \label{fig:rg_vs_ts}
\end{figure}

To investigate if and how aggregates restructure under shear flow, we performed MPCD simulations with heteroaggregates of cubes and cylinders at shear rate $1.36 \times 10^{-4} \tau^{-1}$ and for spheres and rounded cubes at $7.85 \times 10^{-4} \tau^{-1}$ respectively. 
These shear rates represent the highest shear where none of the ten aggregates broke up, instead restructuring of the aggregates was observed. 
The cube MP heteroaggregates restructured under shear flow, as shown by the significant decrease of about $10\%$ in the aggregate radius of gyration, $R_g$, in Fig. \ref{fig:rg_vs_ts}. 
Spheres on the other hand behaved differently, since  they already start as fairly compact aggregates in equilibrium, there was not much possibility to restructure into a less fractal structure (i.e., decrease $R_g$). 
Instead, oscillatory fluctuations of $R_g$ around the mean equilibrium $R_{g,\tau=0}$ were observed, indicating that no large scale restructuring occurred. Those oscillatory fluctuations are caused by the aggregates tumbling in the shear flow. Cylinder and rounded cube aggregates show about $5\%$ decrease of their size.

\begin{figure}
    \centering
    \includegraphics[width=8cm]{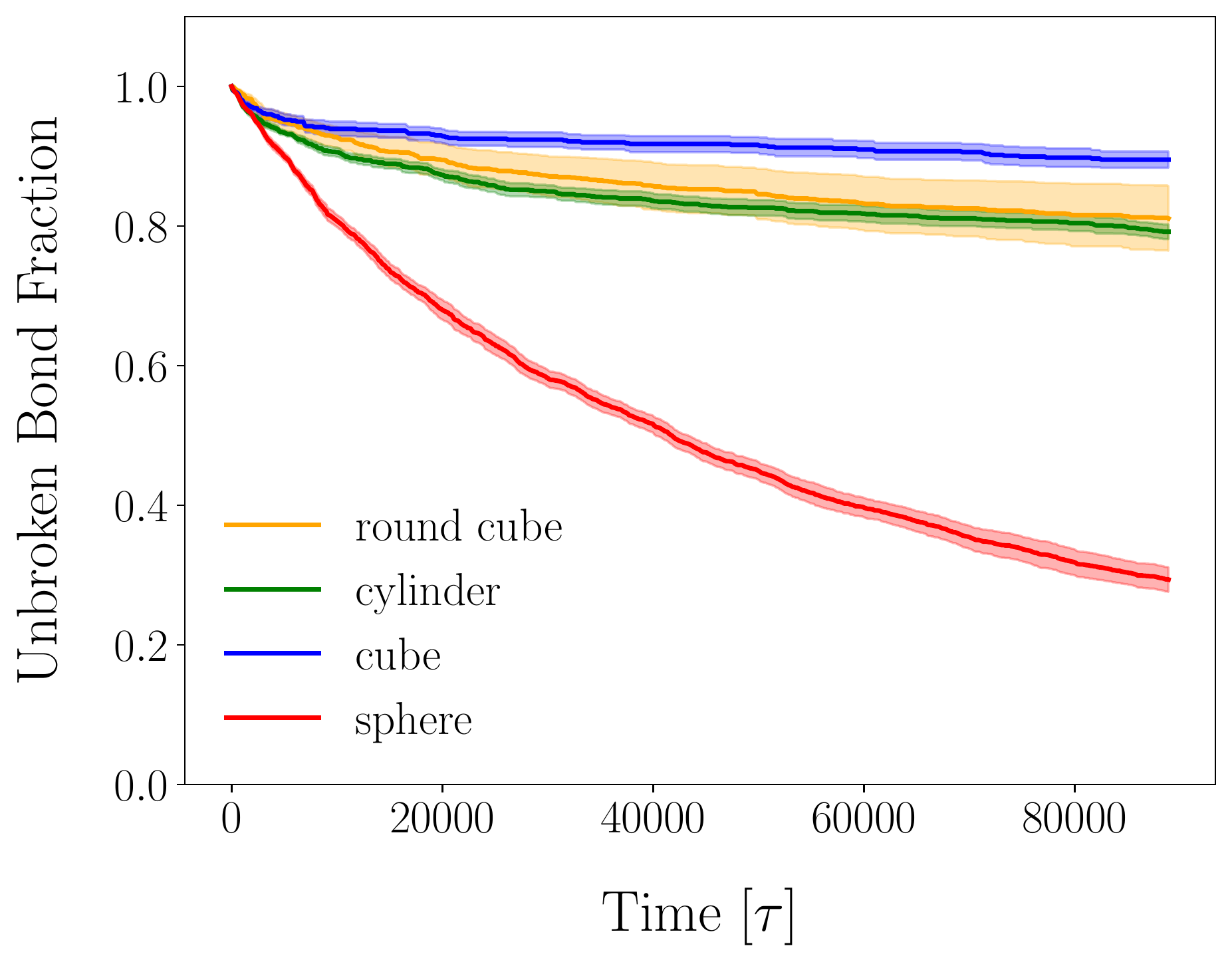}
    \caption{Preserved bond fraction cube and sphere aggregates during restructuring at shear rates $1.36 \times 10^{-4} \tau^{-1}$ and $7.85 \times 10^{-4} \tau^{-1}$ respectively, averaged over 10 aggregates. The shaded area indicates the variance in the averaged data.}
    \label{fig:bonds_vs_ts}
\end{figure}

Fig. \ref{fig:rg_vs_ts} suggests that the overall structure of the more compact sphere MP heteroaggregates were changed significantly less than the fractal cube aggregates, as indicated by changes in $R_g$.
However, on a local, smaller scale the opposite behavior is observed. To measure local rearrangements, 
we defined a preserved bond as a NOM connection that persisted since the equilibrium configuration at $\tau=0$. The resulting fraction of preserved bonds over time during shear flow is plotted in Fig.~\ref{fig:bonds_vs_ts}. 
The fraction of preserved bonds between sphere MPs decayed much quicker compared to all other shapes. 
This suggests that, internally, there was significantly more restructuring of bonds in sphere MP heteroaggregates, where about half of all bonds were changed after approx. $40000\tau$. 
The bonds between two MP cubes were difficult to change due to their higher ACS. More than $90\%$ of initial bonds were preserved after $80000\tau$, with cylinder and rounded cube aggregates in between, where about $80\%$ of bonds were preserved.

Another indication for the difference in internal restructuring can be obtained from the fluctuation of angles formed by bonded triplets in the aggregates. 
Higher fluctuation of angles indicated a stronger internal restructuring for sphere MP aggregates. 
We calculated the standard deviation for each angle that was intact after $20000\tau$, and found that,  
on average, the standard deviation for the angles of sphere aggregates is about 1.4 times higher than cubes. 

Combining the observations made based on $R_g$ and preserved bonds, we can conclude that shear flow was able to easily break the weaker bonds between sphere MPs \textit{without} breaking up the entire heteroaggregate, i.e, minor local rearrangements, whereas the cube heteroaggregates globally restructured into more compact aggregates while preserving about 90\% of all bonds, i.e., a few bonds were broken to allow for entire sections or chunks of the aggregate to move. The rounded cube and cylinder aggregates exhibit restructuring behavior between those two extreme cases. 
While we assume the details of the restructuring to be dependent on the shear rate, the general trends are expected to hold over a range of weak shear rates. 

From the data obtained by the shear MPCD simulation, it was evident that compact heteroaggregates were more resistant against shear flow in general, and consequently, and we can speculate that they may attain larger absolute sizes before they break up due to shear. 
For this reason, we would expect average aggregate sizes of sphere MPs at steady state to be much higher compared to fractal heteroaggregates of cubes.  The stability under shear flow appears to be determined by a combination of compactness and bonded neighbor strength and number of neighbors. 

\section{Conclusions and Outlook \label{sec:conclusion}}
In this work, we provide insights about the impact of shape on the fate of MPs and NOM heteroaggregates in the environment, a parameter which is often overlooked in literature and difficult to quantify.
We found that the MP shape determined the structure of the heteroaggregates, where edges and corners led to more open, fractal clusters with less bonded MP neighbors, but with increased connection strength between each neighboring MP shape. 
More spherical MP shapes led to compact clusters with weaker connections between neighbors. 
Based on the difference in respective fractal dimensions, we speculate that cubes and plates should aggregate faster compared to rounded cubes and spheres.

We also investigated the stability and restructuring of heteroaggregates of different MP shapes under shear. We showed that, in general, a higher number of neighbors and a more compact aggregate (e.g., sphere MPs heteroaggregates) provided more stability against shear breakup than strong connections between neighbors in a fractal aggregate (e.g., cube or plate MPs heteroaggregates). 
Consequently, the critical shear rate for breakup changed with the shape as well, where we found that heteroaggregates of cube and cylinder MPs break up at lower shear rates than sphere heteroaggregates. A notable exception to this trend is the rounded cube MP shape, that exhibited an unexpectedly high stability against breakup under shear. We attribute this to being fairly compact, due to weaker flexible neighbor connections that allow restructuring, but are strong enough to prevent breaking. This result also highlights that measuring a single aggregate property might not be sufficient to describe their non-equilibrium behavior in detail.

This work was focused on systematic variations of aggregate properties and stability under shear as function of shape, therefore  
we have not investigated extremely irregular shapes that can be present in the environment~\citep{wang2021review}.
Future work will include more irregular shapes and heteroaggregation of MP with mixed shapes to investigate the effect of heterogeneity in shapes.

In addition to aggregate breakup and restructuring, sedimentation plays a key role in determining the fate of MPs in water~\citep{domercq2022full}.
It is possible that through fractal dimension, the shape of MPs could have an impact on the sedimentation behaviour of aggregates due to shape drag and other effects of hydrodynamics on settling dynamics~\citep{Gonzalez2004,Mohraz2004,Paul2017}.

Thus far, we have used simple pair interactions, i.e.\, Morse potentials between the constituents in the system. Those pair interactions do not explicitly consider any chemical details beyond generic attraction and repulsion, making the results presented here generically applicable, with the clear limitation of specificity. In the future, (extended) DLVO theories can be used with specific chemistries and solvent conditions to obtain more realistic potentials.
It is also possible to work with more accurate empirical potentials obtained by potential of the mean force (PMF) calculations from experimental results~\citep{Chen2015}.
There is also room for improvement regarding the functional form of the used potentials as well. 
In addition to pair potentials solely based on distance as used in this study, tangential contact potentials and roughness related interactions between colloidal particles can play a significant role in particle behavior~\citep{pantina2005elasticity}.
Because tangential potentials can provide resistance against bending and torsion, they can influence the aggregation process and aggregate behaviour under shear flow. We speculate that addition of roughness and friction will reduce the restructuring of clusters into more compact structures during aggregation, and  will overall result in more fractal aggregates. Improvement of the simple model presented here is left for future work.   

The simple model system used in this study illustrates
the importance of shape of microplastic particles in their aggregation behavior and stability under shear. Overall, edges and corners lead to less stable, more fractal aggregates. The unexpected stability of rounded cube MP aggregates illustrates that for predicting the stability under shear flow, aggregate size, i.e., $d_f$ is not sufficient and more work is needed to fully understand the internal rearrangements of these heterogeneous clusters under shear. 

\section*{Acknowledgements}
 The authors acknowledge the support by the National Science Foundation under Grant DMR 2034496.
This work made use of the Illinois Campus Cluster, a computing resource that was operated by the Illinois Campus Cluster Program (ICCP) in conjunction with the National Center for Supercomputing Applications (NCSA) and which was supported by funds from the University of Illinois at Urbana-Champaign.

\bibliographystyle{elsarticle-harv} 
\bibliography{references}

\end{document}